\documentclass[12pt,preprint]{aastex}

\newcommand{\myemail}{naoyuki@eps.s.u-tokyo.ac.jp}
\newcommand {\Vect}[1]{\mbox{\boldmath $#1$}}

\newcommand {\pdif}[3][]{\frac{\partial^{#1}#2}{\partial#3^{#1}}}

\newcommand {\fr}{\frac}

\def\alfven {Alfv\'{e}n }

\shorttitle{2D MHD simulations of relativistic magnetic reconnection}
\shortauthors{Watanabe \& Yokoyama}

\begin{document}

\title{Two-dimensional MHD simulations of relativistic magnetic reconnection}

\author{Naoyuki Watanabe and Takaaki Yokoyama}
\affil{Department of Earth and Planetary Science, University of Tokyo, 7-3-1, Hongo, Bunkyo-ku, Tokyo, 113-0033, Japan}
\email{\myemail}

\begin{abstract}
It has been recognized that magnetic reconnection process is of great importance in high-energy astrophysics. We develop a new two-dimensional relativistic resistive magnetohydrodynamic (R$^2$MHD) code, and carry out numerical simulations of magnetic reconnection. We find that outflow velocity reaches Alfv\'{e}n velocity in the inflow region, and that higher Alfv\'{e}n velocity provides higher reconnection rate. We also find Lorentz contraction plays an important role in enhancement of reconnection rate.
\end{abstract}


\section{Introduction}

Magnetic reconnection is widely recognized as a very important phenomenon in astrophysics. Over the last decade, it has been recognized that magnetic reconnection processes are very important in high-energy astrophysics. Dissipation of such super strong magnetic fields may play an important role both in global dynamics of the system and as a way to produce high-energy emission. Relativistic magnetic reconnection was proposed as a source of the high-energy emission \citep{1996A&A...311..172L,2002A&A...388L..29K} and as the solution to the $\sigma$-problem \citep{1990ApJ...349..538C,2001ApJ...547..437L,2003ApJ...591..366K,2003MNRAS.339..765L}. Similar models were also developed for the cosmological gamma-ray bursts \citep{2002A&A...387..714D,2002A&A...391.1141D,2001MNRAS.321..177L}. Magnetic reconnection was evoked for explanation of the rapid variability observed in active galactic nuclei \citep{1998MNRAS.299L..15D}. The particle acceleration in the reconnection process was proposed to operate in radio jets \citep{1992A&A...262...26R,2001ApJ...559...96B}. 

Due to the extreme complexity and richness of the possible effects arising in relativistic plasma physics, there is a strong interest for developing computer codes for relativistic magnetohydrodynamics (hereafter RMHD). \cite{1993JCoPh.105..339V} illustrated the implementation on the Riemann problem for MHD. \cite{1996ApJ...463L..71K} then developed a RMHD code, which has been extensively used in relativistic two-dimensional and three-dimensional jet simulations. \cite{1999MNRAS.303..343K} and others developed and tested a Godunov-type code which is a truly multidimensional scheme \citep{2001ApJS..132...83B,2002MNRAS.333..932K}. Recently, \cite{2003A&A...400..397D} presented a third order shock-capturing scheme for three-dimensional RMHD and validated it by several numerical tests. On the other hand, \cite{1998ApJ...495L..63K,1999ApJ...522..727K} extended to general relativistic (GRMHD) effects, and applied it to the jet formation mechanism. \cite{2003ApJ...589..444G} and \cite{2003ApJ...589..458D} also developed GRMHD codes.

Despite magnetic reconnection is recognized as an important process in high-energy astrophysics, there is not a lot of theoretical studies. \cite{1994PhRvL..72..494B} considered kinematics of relativistic reconnection in the Sweet-Parker and Petschek configurations and concluded that due to the Lorentz contraction, the reconnection inflow is significantly enhanced and may approach the speed of light. \cite{2003ApJ...589..893L} confirmed this conclusion for the Sweet-Parker case. \cite{2005MNRAS.358..113L} presented generalization of Sweet-Parker and Petschek reconnection models to the relativistic case, and argued that the reconnection inflow does not approach the speed of light. Particle acceleration in relativistic current sheets was studied both in the test particle approximation \citep{1992A&A...262...26R,2001ApJ...559...96B} and in two-dimensional PIC simulations \citep{2001ApJ...562L..63Z,claus04}. Furthermore, \cite{2005PhRvL..95i5001Z} studied three-dimensional PIC simulations, and suggested the importance of the current-aligned magnetic field for studying the energetics of relativistic current sheet. Meanwhile, there are several RMHD simulations as we write, all these codes, however, are applied to ideal MHD and take no account of resistivity. 

In this paper, we develop a new two-dimensional relativistic resistive MHD (R$^2$MHD) code, and carry out numerical simulations of  two-dimensional relativistic magnetic reconnection.

\section{Simulation model}

The RMHD basic equations are written as follows:
\begin{equation}
\pdif{D}{t}
+\nabla\cdot\left(D{\Vect{v}}\right)
=0
\label{eq:mass}
\end{equation}
\begin{equation}
\pdif{\Vect{R}}{t}
+\nabla\cdot\left[
\left(P
+\frac{B^2+E^2}{8\pi}\right)
{\Vect{I}}
+
\gamma^2(e+P)\frac{\Vect{v}\Vect{v}}{c^2}
-\frac{{\Vect{B}\Vect{B}}
+{\Vect{E}\Vect{E}}}{4\pi}\right]
=0
\label{eq:momentum}
\end{equation}
\begin{equation}
\pdif{\epsilon}{t}+\nabla\cdot
\left[\left\{\gamma^2\left(e+P\right)-D c^2\right\}{\Vect{v}}
+\frac{c}{4\pi}{\Vect{E}}\times{\Vect{B}}\right]
=0
\label{eq:energy}
\end{equation}
\begin{equation}
\pdif{\Vect{B}}{t}
+c\nabla\times{\Vect{E}}
=0
\label{eq:faraday}
\end{equation}
\begin{equation}
\pdif{\Vect{E}}{t}
-c\nabla\times{\Vect{B}}
=-4\pi\Vect{j}
\label{eq:ampere}
\end{equation}
where $c$, $P$, $\Vect{v}$, $\Vect{B}$, $\Vect{E}$, and $\Vect{j}$ are the light speed, proper gas pressure, velocity, magnetic field, electric field, and current density, respectively. $\gamma$ is the Lorentz factor which is defined as $\gamma\equiv[1-(v/c)^2]^{-1/2}$, and $e$ is internal energy given as $e\equiv\rho c^2+P/(\Gamma -1)$, where $\rho$, and $\Gamma$ are the proper mass density, and the specific heat ratio, respectively. $D$, $\Vect{R}$, and $\epsilon$ are defined as, 
\begin{equation}
D=\gamma\rho,
\label{eq:density}
\end{equation}
\begin{equation}
{\Vect{R}}=\gamma^2\left(e+P\right)\frac{\Vect{v}}{c^2}
+\frac{\Vect{E}\times\Vect{B}}{4\pi c},
\label{eq:inertia}
\end{equation}
\begin{equation}
\epsilon=\gamma^2\left(e+P\right)-P-D c^2
+\frac{B^2+E^2}{8\pi}.
\label{eq:epsilon}
\end{equation}
Furthermore, Ohm's law for a relativistic pair plasma under fairly general conditions has the MHD form \citep{1993PhRvL..71.3481B,2003ApJ...589..893L},
\begin{equation}
\gamma\left({\Vect{E}}
+\frac{\Vect{v}}{c}\times\Vect{B}\right)
=\eta\left[{\Vect{j}}+\gamma^2
\left({\Vect{j}}\cdot\frac{\Vect{v}}{c}-\rho_e c\right)\frac{\Vect{v}}{c}\right]
\label{eq:ohm}
\end{equation}
where $\eta$ is the resistivity, and $\rho_e$ is the electron mass density. The second term of right hand side of equation (\ref{eq:ohm}) is the convection current.

When we solve ideal RMHD equations, it is possible to eliminate the electric field using $\Vect{E}=-({\Vect{v}}/c)\times\Vect{B}$. Therefore, past RMHD simulations solved equations (\ref{eq:mass}) - (\ref{eq:faraday}) and evaluate only $D$, $\Vect{R}$, $\epsilon$, and $\Vect{B}$ directly at each step from the equations. In our code, we take into consideration the effect of resistivity, so that we are forced to take another way. We cannot eliminate the electric field, therefore we also solve equation (\ref{eq:ampere}) to evaluate $\Vect{E}$ at each time step. Next, from $D$, $\Vect{R}$, $\epsilon$, $\Vect{B}$, and $\Vect{E}$ obtained by solving equations (\ref{eq:mass}) to (\ref{eq:ampere}), we calculate $\gamma$. For this purpose, we solve an equation for unknown variable $\gamma$,
\begin{equation}
\left[\frac{c\{\Gamma(\gamma^2-1)+1\}}{\Gamma(Dc^2+\epsilon -P_{em})\gamma^2-(\Gamma -1)\gamma Dc^2}\right]^2
|{\Vect{R}}-{\Vect{S}}|^2=1-\frac{1}{\gamma^2}
\label{eq:gamma-eq}
\end{equation}
where ${\Vect{S}}=({\Vect{E}}\times{\Vect{B}})/(4\pi c)$, and $P_{em}=(B^2+E^2)/(8\pi)$, respectively. This equation is obtained by vanishing $\Vect{v}$, $\rho$ and $P$ using equations (\ref{eq:density}) to (\ref{eq:epsilon}). We solved this equation at each cell using the Newton-Raphson iteration method, so that we obtain $\gamma$. We calculate $\Vect{v}$ and $P$ after the iteration, and then we get $\rho$ and $\Vect{j}$ from equations (\ref{eq:density}) and (\ref{eq:ohm}), respectively.

We assume that the evolution is two-dimensional. We take a rectangular computation box with two-dimensional Cartesian coordinates in the $x$-$y$ plane. The medium is assumed to be an inviscid perfect gas. The $z$-component of magnetic field $B_z$, velocity $v_z$, and partial derivative $\partial/\partial z$ are neglected. Electric field $E_z$, and current density $j_z$, however, are included, and there are no $x$ and $y$ components of these variables according to equations (\ref{eq:faraday}) and (\ref{eq:ampere}). Therefore, Ohm's law equation (\ref{eq:ohm}) only has $z$-component, and we can neglect the convection current term. An anomalous resistivity model is assumed, as described later.

The region of the computation box for this study is $-52.2L\leq x \leq 52.2L$, $-151.2L\leq y \leq 151.2L$, where $L=1$ is the thickness of the initial current sheet. Non-uniform grids are used for both $x$- and $y$-directions. The number of grid points is $200\times 416$. The minimum grid sizes are $\Delta x=0.02$ and $\Delta y=0.05$, which are concentrated near the neutral point. 

The light speed $c$ is taken to be unity. The initial proper density outside the current sheet is given as $\rho=\rho_0=1$ in non-dimensional units. We set the initial proper gas pressure outside the current sheet as $P=P_0=1$, and the initial temperature $T=P/\rho =P_0/\rho_0=1$ is uniform everywhere. Magnitude of the magnetic field $B_0=(8\pi P_0/\beta)^{1/2}$ is prescribed by proper gas pressure $P_0$ and the plasma $\beta$. We study several values for $\beta$, but it is $\beta =0.1$ in the typical case. We take a relativistic Harris model as the initial current sheet configuration \citep{2003ApJ...591..366K}, so we give initial conditions as $B_y=B_0\tanh(2x)$, $P=1+[\beta\cosh^2(2x)]^{-1}$, $\rho=1+[\beta\cosh^2(2x)]^{-1}$, $v_x=v_y=0$, and $E_z=\eta(\partial B_y/\partial x)$. Since there is no vertical magnetic field $B_z$, we can write $\Vect{E}=E_z\Vect{\hat{z}}$ by using equations (\ref{eq:faraday}) and (\ref{eq:ampere}). $\eta$ is the resistivity which is defined as:
\begin{eqnarray}
\eta(x,y)=\left\{
\begin{array}{lr}
{\eta_b +\eta_{i0}\left[2(r/r_{\eta})^3-3(r/r_{\eta})^2+1\right]} & {\rm for} \hspace{12pt} r\leq r_{\eta}, \\
\eta_b & {\rm for} \hspace{12pt} r>r_{\eta}, \\
\end{array}
\right.
\end{eqnarray}
where $\eta_b=5.0\times 10^{-3}$ is a uniform resistivity in the computation box, $\eta_{i0}=0.3$ is the amplitude of the anomalous resistivity, $r=\sqrt{x^2+y^2}$ is the distance from the center of the spot (the origin), and $r_{\eta}=0.8$ is the radius of the spot.

\section{Results and Discussion}

\begin{figure}
\includegraphics{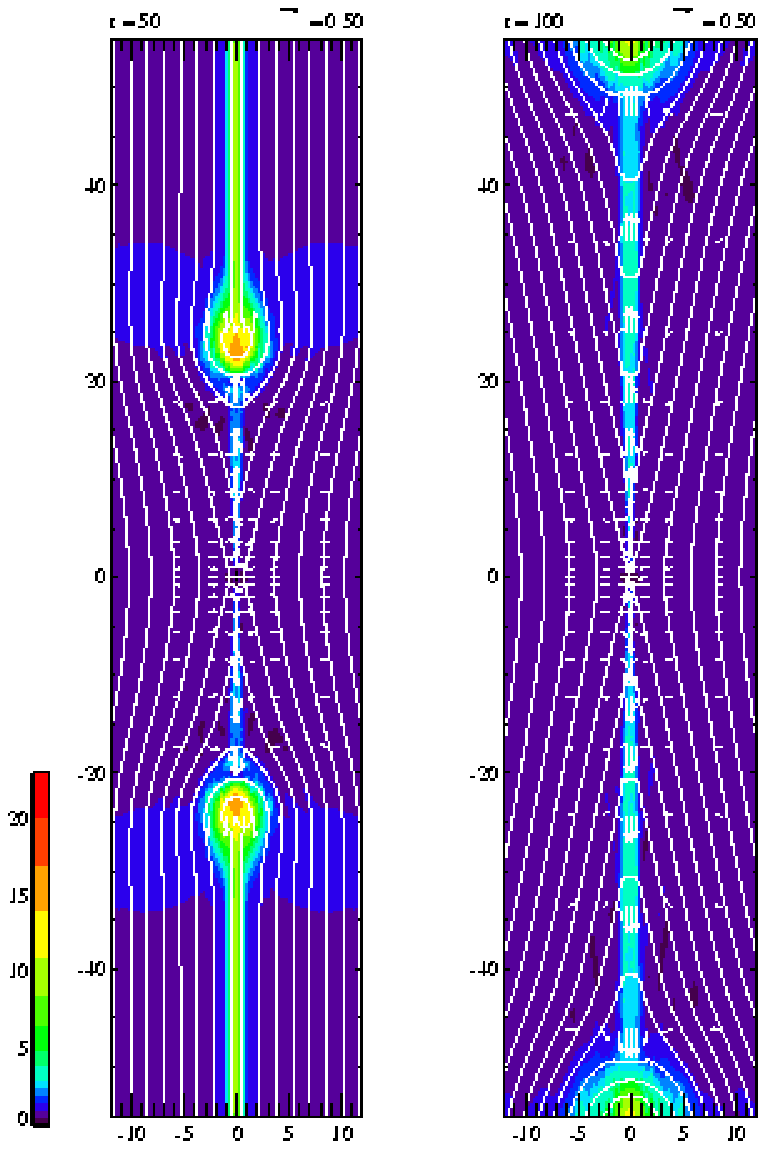}
\caption{\label{fig:2d} Two-dimensional density distributions of typical model ($\beta = 0.1$) at $t=50$ and $t=100$. Solid lines show magnetic field lines, and arrows show velocity vectors.}
\end{figure}

Fig. \ref{fig:2d} shows the density distribution of the typical case ($\beta = 0.1$). Because of the enhanced resistivity around the origin, magnetic reconnection starts at this point. This point evolves to become an X-type neutral point. The reconnected field lines together with the frozen-in plasma are ejected from this X-point to the positive and negative $y$-directions because of the tension force of the reconnected field lines. The velocity of the reconnection outflow $V_{\rm out}\simeq 0.9$ is approximately the \alfven speed of the inflow region ($C_{A0}=0.894$). To complement these outflows, inflows take place from positive and negative $x$-directions of the current sheet. At the boundary between this inflow and the outflow, a shock is formed, emanating from the neutral point.

\begin{figure}
\includegraphics{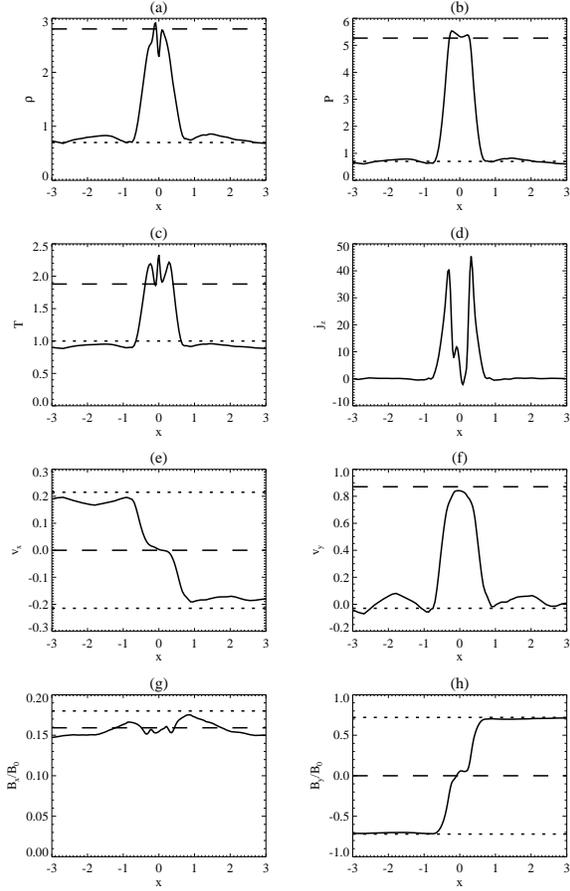}
\caption{\label{fig:1d} One-dimensional distributions of various physical quantities of typical model ($\beta = 0.1$) parallel to the $x$-axis across $y=10$ at $t=100$. Displayed variables are (a) proper density $\rho$, (b) proper gas pressure $P$, (c) temperature $T$, (d) current density $j_z$, (e) $x$-component of velocity $v_x$, (f) $y$-component of velocity $v_y$, (g) $x$-component of magnetic field $B_x/B_0$, and (h) $y$-component of magnetic field $B_y/B_0$. Dashed lines show calculated downstream values, using upstream values those are shown by dotted lines based on the analytical jump conditions.}
\end{figure}

Fig. \ref{fig:1d} shows one-dimensional plots of various physical variables at $t=100$, when the distribution becomes nearly steady state, and along $y=10$, which is well upstream of the plasmoid ejected in the positive $y$-direction. At $x\sim \pm 0.5$, there are strong jumps for several variables. The value of current density $j_z$ becomes large, and $y$-component of magnetic field $B_y$ becomes weak at these jumps. Therefore, we can say these jumps are the slow-mode MHD shocks. We also checked these jump conditions using the arranged model of \citet[shown by dotted and dashed lines]{2005MNRAS.358..113L}. From Fig. \ref{fig:2d} and Fig. \ref{fig:1d}, we obtained $\tan\theta\sim 0.21$ where $\theta$ is the angle between the magnetic field and the shock plane, and this value is close to the inflow velocity at the slow shock (e.g., Fig. \ref{fig:1d}(e)). According to the model of \cite{2005MNRAS.358..113L}, inflow velocity $v_{\rm in}\sim \tan\theta$ in the highly relativistic regime, and our results supports this model.

\begin{figure}
\includegraphics{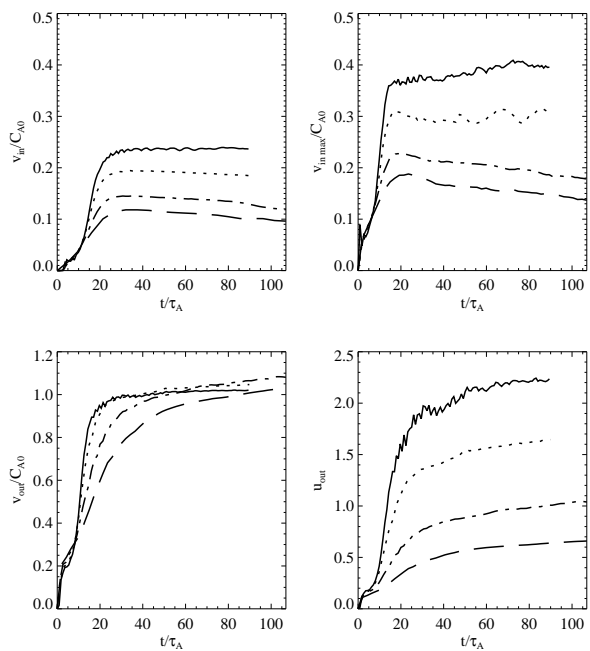}
\caption{\label{fig:para} Plot of time dependence of (a) inflow velocity at $(x,y)=(4,0)$, (b) maximum inflow velocity (almost same as the velocity at the edge of the anomalous resistivity spot $(x,y)\approx(\pm 0.8,0)$), (c) outflow velocity, and (d) outflow 4-velocity for several values of initial plasma $\beta$. Each velocity in (a), (b) and (c) is normalized with \alfven velocity outside the current sheet. Time is normalized with \alfven transit time. Each line shows case for $\beta =0.1$; solid, case for $\beta =0.2$; dotted, case for $\beta =0.5$; dashed and dotted, and case for $\beta =1.0$; dashed.}
\end{figure}

We next studied the dependency on the initial plasma $\beta$. Fig. \ref{fig:para} shows (a) inflow velocity at $x=4$ and $y=0$, (b) maximum inflow velocity, (c) outflow velocity, and (d) outflow 4-velocity as function of time for $\beta =0.1$, 0.2, 0.5, and 1.0 ($C_{A0}=0.894$, 0.816, 0.667, and 0.535, respectively). Velocities are normalized by $C_{A0}$ in (a), (b) and (c), and time is normalized by \alfven transit time $\tau_A=L/C_{A0}$ in all figures. The maximum inflow velocity shown in (b) is almost the same as the inflow velocity at the edge of anomalous resistivity spot ($x\approx\pm 0.8$ and $y=0$). Each line shows the case of a different value of $\beta$. The outflow velocity reaches $C_{A0}$ in all the cases. However, we obtain higher inflow velocity with lower $\beta$ (higher $C_{A0}$). This means higher $C_{A0}$ causes higher reconnection rate $v_{\rm in}/C_{A0}$. In other words, reconnection rate is higher at the relativistic regime.

\begin{figure}
\includegraphics{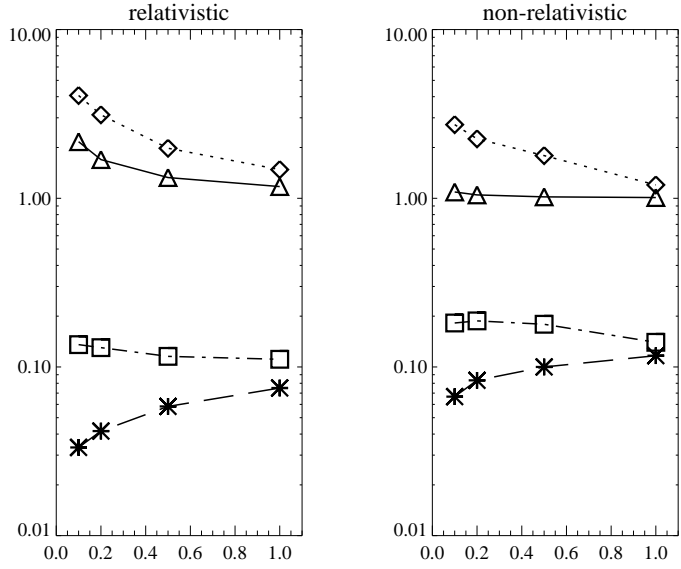}
\caption{\label{fig:ratio} (Left panel) Plot of plasma $\beta$ dependence of $\delta /d$, $\rho_{\rm out}/\rho_{\rm in}$, $(\delta /d)(\rho_{\rm out}/\rho_{\rm in})$, and $\gamma_{\rm out}$ for relativistic reconnection. (Right panel) Plot of plasma $\beta$ dependence of $\delta /d$, $\rho_{\rm out}/\rho_{\rm in}$, $(\delta /d)(\rho_{\rm out}/\rho_{\rm in})$, and $\gamma_{\rm out}$ for non-relativistic reconnection ($P_0 = 10^{-2}$).}
\end{figure}

For a steady state reconnection, we can also express the reconnection rate by using the conservation of the mass at the steady state, $\nabla\cdot (D{\Vect{v}})=\nabla\cdot(\gamma\rho{\Vect{v}})=0$, so that we obtain a following equation:
\begin{equation}
\fr{v_{\rm in}}{v_{\rm out}}\approx\fr{\delta}{d}
\fr{\rho_{\rm out}}{\rho_{\rm in}}\gamma_{\rm out}
\label{eq:rate}
\end{equation}
where $\rho_{\rm out}$ and $\rho_{\rm in}$ are proper density of the inflow and the outflow region, and $\gamma_{\rm out}$ is the Lorentz factor of the outflow velocity, respectively. $\delta$ and $d$ are evaluated by $\delta/d=\tan\theta$, where $\theta$ is the angle between the $y$-axis and the slow-shock plane. We consider the Lorentz factor of the inflow velocity $\gamma_{\rm in}\sim 1$. Fig. \ref{fig:ratio} shows the dependency of $\delta /d$, $\rho_{\rm out}/\rho_{\rm in}$, $(\delta /d)(\rho_{\rm out}/\rho_{\rm in})$, and $\gamma_{\rm out}$ to initial plasma $\beta$ under the relativistic regime ($P_0 = 1.0$), and the non-relativistic regime ($P_0 = 10^{-2}$). From these panels, we can see the similar behaviors of ratios $\delta /d$ and $\rho_{\rm out}/\rho_{\rm in}$ in the both regimes. Furthermore, $(\delta /d)(\rho_{\rm out}/\rho_{\rm in})\sim 0.11 - 0.14$ under the relativistic regime, while $(\delta /d)(\rho_{\rm out}/\rho_{\rm in})\sim 0.14 - 0.18$ under the non-relativistic regime. Therefore the product of the two terms are roughly constant in both regimes, and we can say that the effect of the Lorentz factor, namely, the Lorentz contraction is an important factor to determine the reconnection rate under the relativistic regime.

Let us summarize this paper. The motivation of this study is to investigate relativistic effects of magnetic reconnection to apply for high-energy phenomena. For this purpose, what we have done were; (i) to develop a new resistive relativistic MHD code, and (ii) to do numerical simulations for relativistic magnetic reconnection. From our study, we obtain that outflow velocity become close to the light speed, and due to the high inflow velocity, high reconnection rate is obtained. For the enhancement of the reconnection rate, we find that the effect of the Lorentz contraction is significant which is suggested by \cite{1994PhRvL..72..494B}. However, our results also supports the suggestions of \cite{2005MNRAS.358..113L}, specially in the physics of jump conditions at the slow shocks. We recognize that simulations at the ultra-relativistic regime are required, so we would like to report these results in future.

\acknowledgments

The authors are greatful to Prof. T. Terasawa, Prof. M. Hoshino, Prof. S. Koide, Prof. S. Shibata, Dr. M. Kino and Dr. S. Zenitani for fruitful discussions. This work was supported by facilities of JAXA.

\bibliographystyle{apj}

\clearpage

\clearpage


\clearpage

\clearpage

\end{document}